# Damage threshold evaluation of thin metallic films exposed to femtosecond laser pulses: the role of material thickness

G. D. Tsibidis [1,2,♣], D. Mansour [1], and E. Stratakis [1,3,♦]

[1] *Institute of Electronic Structure and Laser (IESL), Foundation for Research and Technology (FORTH), N. Plastira 100, Vassilika Vouton, 70013, Heraklion, Crete, Greece*

[2] *Department of Materials Science and Technology, University of Crete, 71003 Heraklion, Greece*

[3] *Department of Physics, University of Crete, 71003 Heraklion, Greece*

Email: [♣]tsibidis@iesl.forth.gr; [♦]stratak@iesl.forth.gr

**Abstract**

The employment of femtosecond pulsed lasers has received significant attention due to its capability to facilitate fabrication of precise patterns at the micro- and nano- lengths scales. A key issue for efficient material processing is the accurate determination of the damage threshold that is associated with the laser peak fluence at which minimal damage occurs on the surface of the irradiated solid. Despite a wealth of previous reports that focused on the evaluation of the laser conditions that lead to the onset of damage, the investigation of both the optical and thermal response of thin films of sizes comparable to the optical penetration depth is still an unexplored area. In this report, a detailed theoretical analysis of the impact of various parameters such as the photon energies and material thickness on the damage threshold for various metals (Au, Ag, Cu, Al, Ni, Ti, Cr, Stainless Steel) is investigated. A multiscale physical model is used that correlates the energy absorption, electron excitation, relaxation processes and minimal surface modification. The satisfactory agreement of the theoretical model with some experimental results indicates that the damage threshold evaluation method could represent a systematic approach towards designing efficient laser-based fabrication systems and optimizing the processing outcome for various applications.

# I. INTRODUCTION

Laser surface processing has emerged as a fast, chemical- free technique for surface patterning and functionalization. In particular, the use of femtosecond (fs) pulsed laser sources for material processing has received considerable attention due to the important technological applications [1-7]. The direct correlation of the response of irradiated solids, the exploration of laser-driven phenomena and the induced surface topographies with potential applications require a precise knowledge of the fundamental mechanisms that characterise laser-matter interaction. More specifically, a systematic analysis of the ultrafast phenomena that occur ranging from electron excitation to thermalisation of the electron system and transfer of their energy to the lattice is crucial to provide a detailed description of the multiscale processes; a thorough knowledge of these physical processes is expected to enhance the capability to control laser energy towards fabricating application-based topographies. Thus, the elucidation of the complex physical mechanisms appears to be very critical both from a fundamental and application point of view. To this end, a plethora of consistent methodologies have been developed to explore experimentally and theoretically the multiscale phenomena [8-20].

A key issue for efficient material processing is the accurate determination of the damage threshold which is defined as the minimum laser peak fluence for which an observable modification occurs on the surface of the irradiated solid. In principle, this minimal surface damage is a related to a phase transition (i.e. material melting) and mass displacement [8, 21]. Various standard and accurate methods for the estimation of the damage threshold have been developed and presented in previous reports [22-25]. From a theoretical point of view, the prediction of the damage threshold was performed considering a thermal criterion (i.e. the lattice temperature exceeds the melting point) [8, 21, 26, 27] through the use of the classical Two Temperature Model (TTM) [28] which describes the electron-phonon temperature dynamics and relaxation process [29]. Alternative studies also included the employment of atomistic continuum models with the combination of Molecular Dynamics and TTM [17, 29].

Nevertheless, most of the current research focused on the analysis of data aiming to describe conditions leading to minimal surface modification on bulk materials. On the other hand, due to the increasing interest in patterning of thin solid films (of sizes comparable to the optical penetration depth) for various applications related to optics, healthcare, sensing, environment, energy [30-40], a detailed investigation of the ultrafast dynamics for such materials is imperative. Although results have been reported in some previous works for a variety of materials [41-45], a combined theoretical exploration of the multiscale phenomena that take place as a function of the material thickness is still elusive. Results from various experiments have shown that the optical properties of thin materials deviate from those of bulk solids as the thickness decreases to sizes comparable to the optical penetration depth [44]. These results indicate that the level of the absorbed energy will also differ from whether the material is treated as a bulk solid which, in turn, is expected to be reflected on the damage threshold. Similarly, thinner films appear to inhibit the electron diffusion which delays the electron-phonon coupling and leads to lower damage thresholds [41, 43, 44, 46]. On the other hand, it is important to evaluate the opto/thermal response of materials that are classified as noble (such as copper, gold, silver) or transition materials (such as nickel, titanium); this is due to the fact that the distinctly different electron distribution of those materials [47] appears to affect the optical and thermal properties and therefore, the damage threshold is expected to vary at decreasing thickness.

The elucidation of the aforementioned issues through the development of a comprehensive theoretical framework is of paramount importance not only to understand the complex physical mechanisms of laser-matter interactions for thin films but also to associate the resulting thermal effects with targeted patterning strategies. To this end, in this work, we present a theoretical model (Section II) that can be used to predict the damage threshold for various metals (Au, Ag, Cu, Al, Ni, Ti, Cr, Stainless Steel) as a function of the material thickness. A library of results is derived through the evaluation of variation of the dielectric parameters as a function of the material thickness through the application of a 'multiple reflection' algorithm [48]. A dynamic variation of the optical properties within the pulse duration is also considered which provides appropriate corrections to the excitation level reached by the electron system. Relaxation processes are described through the employment of a TTM and a melting-point-based thermal criterion is used to determine the damage threshold. The investigation has been performed for laser sources at two different spectral regions (515 nm and 1026 nm) to reveal the role of the photon energy in the damage threshold and optical parameters. To validate the model, experimental results for three materials (nickel, chromium, and gold) are tested against theoretical results (Section III). A systematic analysis of the results is illustrated in Section IV while concluding remarks follow in Section V.

## II. THEORETICAL MODEL

To describe the damage induced on the material following irradiation with fs pulses, a theoretical framework is employed to explore the excitation and thermal response of a double-layered structure (thin metal film on a dielectric material). The simulation algorithm is based on the use of a Two Temperature Model (TTM) that represents the standard approach to evaluate the dynamics of electron excitation and relaxation processes in solids [28]. In this work, for the sake of simplicity, an 1D-TTM is used to describe the thermal effects due to heating of the thin films with laser pulses of wavelength $\lambda_L$=515 nm and 1026 nm and pulse duration equal to $\tau_p$=170 fs. It is noted that the solution of the TTM in 1D (along the energy propagation direction) is a standard approach used to describe thermal effects after irradiation of solids with femtosecond laser pulses [49, 50]. This is a reasonable choice assuming that, in principle, the laser spot radius is much larger (some tens of micrometers) than the thickness of the irradiated solid; therefore, the laser energy distribution along the lateral direction is considered uniform which means that the heat conduction along that direction is practically infinitesimal. Due to the presence of the substrate, the following set of rate equations is employed [51]

$$
\begin{aligned}
C_e^{(m)} \frac{\partial T_e^{(m)}}{\partial t} &= \frac{\partial}{\partial z}\left(k_e^{(m)} \frac{\partial T_e^{(m)}}{\partial z}\right) - G_{eL}^{(m)}\left(T_e^{(m)} - T_L^{(m)}\right) + S^{(m)} \\
C_L^{(m)} \frac{\partial T_L^{(m)}}{\partial t} &= \frac{\partial}{\partial z}\left(k_L^{(m)} \frac{\partial T_L^{(m)}}{\partial z}\right) + G_{eL}^{(m)}\left(T_e^{(m)} - T_L^{(m)}\right) \\
C_e^{(S)} \frac{\partial T_e^{(S)}}{\partial t} &= \frac{\partial}{\partial z}\left(k_e^{(S)} \frac{\partial T_e^{(S)}}{\partial z}\right) - G_{eL}^{(S)}\left(T_e^{(S)} - T_L^{(S)}\right) + S^{(S)} \\
C_L^{(S)} \frac{\partial T_L^{(S)}}{\partial t} &= \frac{\partial}{\partial z}\left(k_L^{(S)} \frac{\partial T_L^{(S)}}{\partial z}\right) + G_{eL}^{(S)}\left(T_e^{(S)} - T_L^{(S)}\right)
\end{aligned}
\quad (1)
$$

where the subscript '$m$' (or '$S$') indicates the thin film (or substrate). In Eqs.1, $T_e^{(m)}$ and $T_L^{(m)}$ stand for the electron and lattice temperatures, respectively, of the upper layer. The thermophysical properties of the metal such as the electron $C_e^{(m)}$ (or lattice $C_L^{(m)}$) volumetric heat capacities, electron $k_e^{(m)}$ $\left(= k_{e0}^{(m)} \frac{B_e T_e^{(m)}}{A_e\left(T_e^{(m)}\right)^2 + B_e T_L^{(m)}}\right)$ heat conductivity, the electron-phonon coupling strengths $G_{eL}^{(m)}$, $A_e$, $B_e$ and other model parameters that appear in the first two equations are listed in Table 1 (note $A_g$=194 Jm$^{-3}$K$^{-1}$ for Cr). It is emphasised that as heat conduction in metals is, in principle, due to electrons, the thermal conductivity of the lattice is substantially smaller than that of the electron system. To include this large difference, $k_L^{(m)}$ is normally taken either equal to zero ($k_L^{(m)}$=0) [44, 50] or equal to a very small value compared to the electron heat conductivity ($k_L^{(m)}$=0.01$k_e^{(m)}$ is an expression that has been used in other reports [52, 53]). In this work and without loss of generality, we adopt the latter assumption.

The quantity $S^{(m)}$ represents the source term that accounts for the energy that the laser source provides to the metal surface which is sufficient to generate excited carriers on the thin film. As the purpose of the present investigation is to reveal the impact of optically excited *thin* films, it is important to take to into account the following processes: (i) a portion of the energy is absorbed from the material while part of the laser energy is transmitted into the substrate, (ii) the reflectivity and transmissivity of the irradiated material are influenced by a *multiple reflection process* between the two interfaces (air/metal and metal/substrate), (iii) the transmitted energy into the substrate is not sufficiently high to generate excited carriers and therefore, the third equation of Eqs.1 can be removed while the fourth can be simplified by $C_L^{(S)} \frac{\partial T_L^{(S)}}{\partial t} = \frac{\partial}{\partial z}\left(k_L^{(S)} \frac{\partial T_L^{(S)}}{\partial z}\right)$ where $T_L^{(S)}$, $C_L^{(S)}$, $k_L^{(S)}$ stands for the substrate temperature, volumetric heat capacity and heat conductivity, respectively. The expression for the source term $S^{(m)}$ which is used to excite a metallic surface of thickness $d$ is given from the following formula [44]

$$S^{(m)} = \frac{(1-R-T)\sqrt{4\log(2)}F}{\sqrt{\pi}\tau_p(\alpha^{-1}+L_b)} \exp\left(-4\log(2)\left(\frac{t-3t_p}{t_p}\right)^2\right) \frac{1}{(1-\exp(-d/(\alpha^{-1}+L_b))} \quad (2)$$

where $R$ and $T$ stand for the reflectivity and transmissivity, respectively, $L_b$ corresponds to the ballistic length, $\alpha$ is the absorption coefficient that is wavelength dependent, and $F$ is the peak fluence of the laser beam. The ballistic transport is also included in the expression as it has been demonstrated that it plays significant role in the response of the material [44]. Special attention is required for the ballistic length as in previous works, it has been reported that for bulk materials, $L_b$ in *s/p*-band metals are comparable ($L_b^{(Au)}$=100 nm, $L_b^{(Ag)}$=142 nm, $L_b^{(Cu)}$=70 nm, $L_b^{(Al)}$=46 nm [44]) while for the *d*-band metals such as Ni, Ti, Cr, stainless steel, it is of the same order as their optical penetration depth [44]. It is noted that the above values of the ballistic length were used for $\lambda_L$~ 400-500 nm pulses [41, 44]. The photon energy of the laser beam is expected to influence the ballistic length of the electrons [54] and therefore, the use of the above values for $L_b$ at higher wavelengths (i.e. 1026 nm) could be questionable. Nevertheless, in the absence of appropriate values for $L_b$ at higher photon energies in bibliography, the above values for the ballistic length are used.

The calculation of $R$ and $T$ and the absorbance $A=1-R-T$ are derived through the use of the multiple reflection theory [48]. Thus, the following expressions are employed to calculate the optical properties for a thin film on a substrate (for a *p*-polarised beam)

$$R = |r_{dl}|^2, \quad T = |t_{dl}|^2 \tilde{N}_S, \quad r_{dl} = \frac{r_{am}+r_{mS}e^{2\beta j}}{1+r_{am}+r_{mS}e^{2\beta j}}, \quad t_{dl} = \frac{t_{am}t_{mS}e^{\beta j}}{1+r_{am}+r_{mS}e^{2\beta j}}, \quad \beta = 2\pi d/\lambda_L$$
(3)

$$r_{CD} = \frac{\tilde{N}_D - \tilde{N}_C}{\tilde{N}_D + \tilde{N}_C}, \quad t_{CD} = \frac{2\tilde{N}_C}{\tilde{N}_D + \tilde{N}_C} \quad (4)$$

where the indices $C=a,m$ and $D=m,S$ characterise each material ('a', 'm', 'S' stand for 'air', 'metal', 'substrate', respectively). The complex refractive indices of the materials such as air, metal and substrate are denoted with $\tilde{N}_a = 1$, $\tilde{N}_m = Re(\tilde{N}_m) + Im(\tilde{N}_m)j$, $\tilde{N}_S = Re(\tilde{N}_s)$, respectively. Given that soda lime silica glass is used as a substrate material, $Re(\widetilde{N}_s)(\lambda_L = 1026\ nm) \cong 1.5134$ and $Re(\widetilde{N}_s)(\lambda_L = 515\ nm) \cong 1.5271$ [55] are the refractive indices of the substrate at the two laser wavelengths used in this work. It is noted that a Drude-Lorentz model is used to obtain the dielectric function for each metal based on the analysis by Rakic *et* al. (where both interband and intraband transitions are assumed) [56]. As the optical parameters of an excited material does not remain constant during the excitation process [18], to introduce the transient change, a temporally varying expression of the dielectric function is provided by including a temperature dependence on the reciprocal of the electron relaxation time $\tau_e$ (i.e. $\tau_e = \left[A_e\left(T_e^{(m)}\right)^2 + B_e T_L^{(m)}\right]^{-1}$) [57]. The values of the refractive indices of the metals in this study (at 300 K are given in Table 1).

The volumetric heat capacity of soda lime silica glass is $C_L^{(S)} = 2.1\times10^6$ Jm$^{-3}$K$^{-1}$ while the heat conductivity of soda lime silica glass is equal to $k_L^{(S)} = 1.06$ Wm$^{-1}$K$^{-1}$. The set of equations Eqs.1-4 are solved by using an iterative Crank-Nicolson scheme based on a finite-difference method. It is assumed that the system is in thermal equilibrium at *t*=0 and, therefore, $T_e^{(m)}(z, t = 0) = T_L^{(m)}(z, t = 0) = 300$ K. A thick substrate is considered (i.e. $k_L^{(S)} \frac{\partial T_L^{(S)}}{\partial z} = 0$) while adiabatic conditions are applied on the surface of the metallic surface (i.e. $k_e^{(m)} \frac{\partial T_e^{(m)}}{\partial z} = 0$). Finally, the following boundary conditions are considered on the interface between the top layer and the substrate: $k_L^{(m)} \frac{\partial T_L^{(m)}}{\partial z} = k_L^{(S)} \frac{\partial T_L^{(S)}}{\partial z}, k_e^{(m)} \frac{\partial T_e^{(m)}}{\partial z} = 0, T_L^{(m)} = T_L^{(S)}$.

| | Material | | | | | | | |
|---|---|---|---|---|---|---|---|---|
| **Parameter** | **Au** | **Ag** | **Cu** | **Al** | **Ni** | **Ti** | **Cr** | **Steel (100Cr6)** |
| $\widetilde{N}_m$ | DL [56] | DL [56] | DL [56] | DL [56] | DL [56] | DL [56] | DL [56] | DL [57] |
| $G_{eL}^{(m)}$ [Wm$^{-3}$K$^{-1}$] | Ab-Initio [47] | Ab-Initio [47] | Ab-Initio [47] | Ab-Initio [47] | Ab-Initio [47] | Ab-Initio [47] | $42\times10^{16}$ [44] | Ab-Initio [58] |
| $C_e^{(m)}$ [Jm$^{-3}$K$^{-1}$] | Ab-Initio [47] | Ab-Initio [47] | Ab-Initio [47] | Ab-Initio [47] | Ab-Initio [47] | Ab-Initio [47] | $A_g T_e^{(m)}$ [44] | Ab-Initio [58] |
| $C_L^{(m)}$ [$\times10^6$ Jm$^{-3}$K$^{-1}$] | 2.48 [44] | 2.5 [53] | 3.3 [44] | 2.4 [59] | 4.3 [44] | 2.35 [60] | 3.3 [44] | 3.27 [58] |
| $k_{e0}^{(m)}$ [Wm$^{-1}$K$^{-1}$] | 318 [44] | 428 [53] | 401 [44] | 235 [59] | 90 [44] | 21.9 [60] | 93.9 [44] | 46.6 [58] |
| $A_e$ [$\times10^7$ s$^{-1}$K$^{-2}$] | 1.18 [53] | 0.932 [53] | 1.28 [53] | 0.376 [59] | 0.59 [53] | 1 [60] | 7.9 [52] | 0.98 [58] |
| $B_e$ [$\times10^{11}$ s$^{-1}$K$^{-1}$] | 1.25 [53] | 1.02 [53] | 1.23 [53] | 3.9 [59] | 1.4 [53] | 1.5 [60] | 13.4 [52] | 2.8 [58] |
| $T_{melt}$ [K] | 1337 [61] | 1234 [61] | 1357 [61] | 933 [61] | 1728 [61] | 1941 [61] | 2180 [61] | 1811 [61] |

Table 1: Optical and thermophysical properties of materials (*DL* stands for Drude-Lorentz model).

It has to be emphasised that a more rigorous approach would also require the investigation of the impact of the electron (in the metal) –phonon (in the dielectric) scattering (EM-PD) heat transfer across metal-dielectric surfaces. More specifically, in previous reports, it has been shown that, especially for very thin films, the impact of EM-PD coupling increases at decreasing thicknesses and this phenomenon leads to remarkable variation in the relaxation process [62]. Thus, special emphasis on the role of the thermal resistance (related to the reciprocal of the EM-PD coupling) should be given. Nevertheless, in those studies, the use of approximate expressions both for the thermophysical and the optical parameters of the irradiated metals as well as the neglect of the relation of the optical parameters variation with thickness (i.e. application of the multiple reflection theory) do not allow a direct and consistent interpretation with experimental data. Furthermore, in those studies, a fitting approach through the use of appropriate experimental protocols was performed to estimate the thermal resistance. Thus, despite a non-vanishing EM-PD coupling, the evaluation of the associated thermal resistance is not straightforward. In this study, the incorporation of the role of thermal resistance on the thermal effects on the irradiated material has not been used; however, the evaluation of the damage threshold and comparison with experimental data are expected to function as a test for the need of a more complex (but, anyway, more consistent with the application of all underlying physical mechanisms).

## III. EXPERIMENTAL PROTOCOL

To determine the damage threshold, $F_{thr}$, of thin metal films we used a single series of single shot experiments on thin films of various thicknesses. More specifically, Cr metal targets of various thicknesses (10 nm, 20 nm, 40 nm, 100 nm, and 300 nm) and Ni thin films of various thicknesses (10 nm, 20 nm, 40 nm, and 100 nm) were deposited on soda lime silica glass plates of 1 mm thickness by e-beam evaporation [63]. In regard to the surface conditions, the roughness depends on the substrate (a microscope slide glasses made of soda lime glass). The deviation of thickness to the nominal values were <6% (for Cr), and <5% (for Nickel). We have included this information in the revised version of the manuscript. The following experimental setup was used for the calculation of the damage threshold of Ni and Chromium Cr thin films: linearly polarized, Gaussian laser pulses of wavelength of $\lambda_L$= 1026 nm and pulse duration $\tau_p = 170$ fs were used for the irradiation of the samples. The laser beam is steered with the help of mirrors, and it was focused, at normal incidence onto the target using a converging lens with focal length equal to $f = 200$ mm, resulting in a focal spot of waist radius $w = 28.7$μm. In all cases, the samples were placed inside a vacuum chamber, to avoid oxidization of the metals, with constant pressure $p = 7 \times 10^{-2}$ mbar, the chamber had a window of a fused silica glass plate in order to be transparent to the laser beam which processed the thin metal films.

## IV. RESULTS AND DISCUSSION

The theoretical model presented in Section II is aimed to describe the ultrafast dynamics and the response of the electron and lattice subsystems. The objective of the current study is to reveal the impact of the material thickness, firstly, on the absorbed energy and, secondly, on the thermal response of the material.

The ultrafast dynamics following irradiation of two different materials, one *s/p*-band metal (Au) and a *d*-band metal (Ni) with single laser pulses of $\lambda_L = 515$ nm was explored while similar conclusions can be deduced for the rest of the metals in this study and at different laser wavelengths. The analysis is, firstly, focused on moderate fluences that are not sufficient to cause damage to the material which are capable, though, to reveal the distinctly different thermal response of the two materials. Our simulation results at $F$=10 mJ/cm² for two distinctly different thicknesses, 20 nm and 400 nm, illustrate remarkably contrasting behaviour of the electron and lattice temperatures *(on the surface)* for the two metals (Figs.1-2). More specifically, for both metals, a decrease of the material thickness leads to a smaller depth in which the electrons can diffuse. It is known that the electron subsystem loses energy (cools down) through two competing effects, diffusion and electron-phonon scattering. Due to the fact that electron diffusion is inhibited from the small thickness of the film, hot electrons remain confined in a small volume and they lose energy, in principle, through electron-phonon interaction. Therefore, the electron temperature decrease is not as rapid as in the case of a bulk material for both Au and Ni (Fig.1) and the diffusive part in the first equation of Eqs.1 could be ignored. Thus, for small thicknesses, the phonon subsystem will interact with an electron system, which are highly energetic and attain large temperatures.

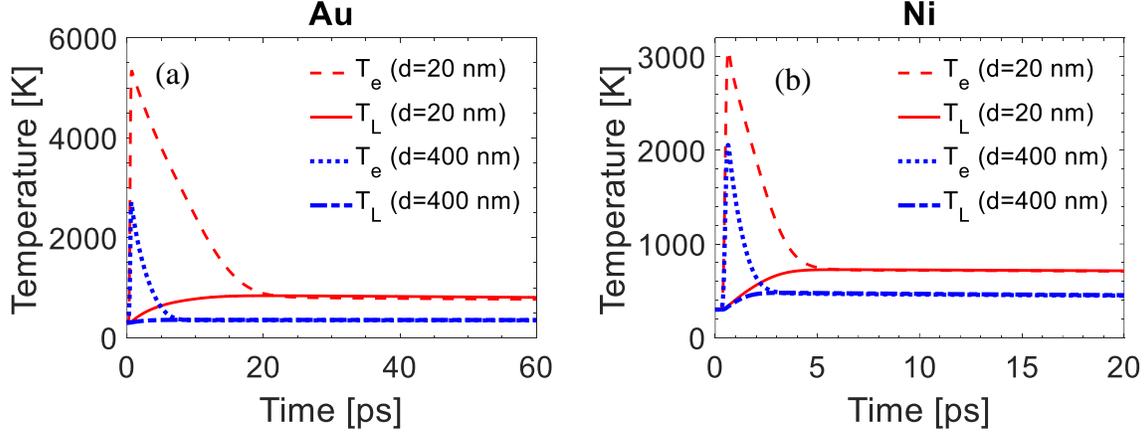

Figure 1: Electron and lattice temperature evolution for (a) Au (20 nm and 400 nm), (b) Ni films (20 nm and 400 nm) ($\lambda_L$=515 nm). Simulation results indicate thermal variation on the surface of the solid.

This means that upon relaxation, the produced lattice temperature will be higher than the maximum lattice temperature obtained from the phonon system in thicker materials. Similar results have been reported in previous works as well, however, in those studies approximate values for the thermophysical properties, energy absorption were assumed while the influence of the thickness on the optical properties was ignored [44]; thus, while from a qualitative point of view, the derived results were adequate to present a picture of the ultrafast dynamics, fitting of data were required to correlate the theoretical predictions with experimental observables such as surface modification. Furthermore, to interpret the increase in the maximum $T_e$ value for thin films and compare it with simulations prediction for thicker films, we need to look at the interplay of the absorbance and the term $\frac{1}{(1-\exp{(-d/(\alpha^{-1}+L_b))})}$ in Eq.2. Our results (Figs.3,5) indicate that for Au and Ni at $\lambda_L = 515$ nm, the absorbance drops at decreasing thickness which would rather lead to an increase of the (maximum) electron temperature. On the other hand, simulations in a previous report (considering the same source term but the same absorbance value for small (20 nm and bulk Au) showed that identical maximum electron temperatures can be obtained only if higher fluences are used for bulk materials [44]. This suggests that irradiation of Au at the same fluence and absorbance yields higher maximum $T_e$ on the surface of the solid. A closer look of the contribution of the term $\frac{1}{(1-\exp{(-d/(\alpha^{-1}+L_b))})}$, though, shows that this quantity plays a crucial role in the magnitude of the source term. As $\frac{1}{(1-\exp{(-d/(\alpha^{-1}+L_b))})}$ increases at smaller thickness, the absorbed energy increases which, in turn, leads to the behaviour illustrated in Fig.1.

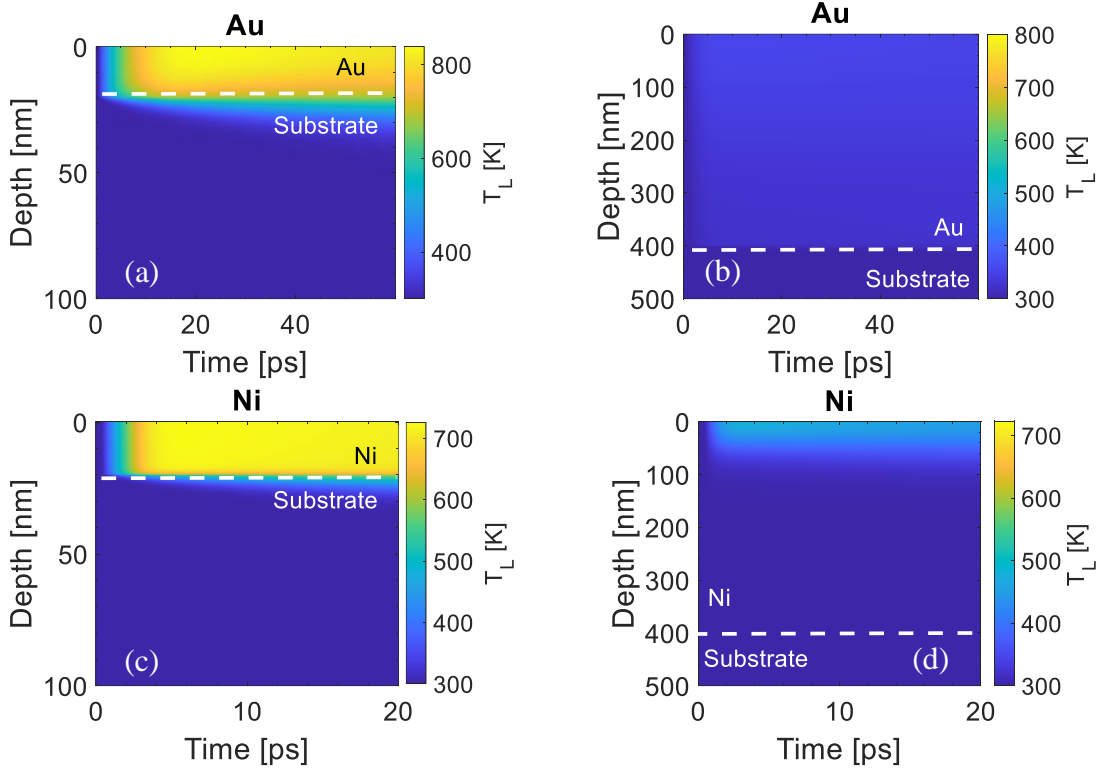

Figure 2: Spatio-temporal lattice temperature evolution for (a) Au (20 nm), (b) Au (400 nm), (c) Ni (20 nm), (d) Ni (400 nm) ($\lambda_L$=515 nm). The *dashed* white line indicates the interface between the metal and the substrate.

Another interesting conclusion from our analysis is that due to the delay in the drop of the electron temperature value, the relaxation of the system will occur at longer times compared with what happens in thicker (or bulk) materials. In Fig.2, the spatiotemporal profile of the lattice temperature field is also illustrated that also shows the enhanced temperature obtained at smaller metal thicknesses. A comparison of the dynamics and the temperature evolution for the two metals manifests a faster relaxation process for Ni than for Au and this monotonicity holds also for thinner films (Fig.1). This is due to the fact that the electron-phonon coupling for Ni is substantially larger at small temperatures [47] (in principle, $G_{eL}^{(m)}$ exhibits an opposite monotonicity at increasing electron temperature compared to Au) which accelerates the relaxation process. It is noted that lattice temperature in Au spreads faster in a larger volume (Fig.2b) due to both the larger ballistic electron transport distance and electron diffusion compared to Ni (Fig.2d); this implies an enhanced presence of hot electrons in Au in deeper regions that interact strongly with the phonon system and therefore, larger temperatures are generated inside higher volumes for Au. The above observation is more pronounced for thicker films (Fig2b,d); results do not show a substantial change for thinner films (Fig.2a,c) as the preferential electron diffusion for Au is prevented from the thickness of the material. This behaviour is expected assuming the imposed boundary condition $k_e^{(m)} \frac{\partial T_e^{(m)}}{\partial z} = 0$ and considering that the transmitted beam is not sufficient to excite the substrate.

The aforementioned investigation showed that for both Au and Ni and despite their different thermophysical properties and electronic structure, the decrease of the metal thickness leads to a delay in the relaxation process due to the inhibited electron diffusion. In particular, even for materials of distinctly different electronic structure and electron-phonon coupling dependence on $T_e$, there is a delay in the time of equilibration between the electron and lattice systems. A similar procedure can be followed to provide a quantification of the relaxation time at various thicknesses for the rest of the materials, however, this is outside the scope of the present study. On the other hand, it is very important to reveal the impact of the thickness on the thermal response not only through the inhibited electron diffusion (at smaller thicknesses) and differences in the electron-phonon coupling but also through the variation of the optical response and therefore energy absorption at various thicknesses. Simulations results indicate that there is distinct decrease of the reflectivity for all metals at both laser wavelengths as the thickness size lowers. More specifically, the use of the multiple reflection theory and Eqs.3-4 predict substantially smaller reflectivity values for thicknesses close to the penetration depth (~15-20 nm) compared to the values for thicker (or bulk) materials (see *left* column in Figs.3-5). On the other hand, a substantially smaller variation of the absorbance is derived for thin and thicker films (*middle* column in Figs.3-5). These results demonstrate that the enhanced transmissivity for thinner films (calculated through the expression *T=1-A-R*) is responsible for the drop of the reflectivity. These theoretical predictions have been confirmed experimentally [42].

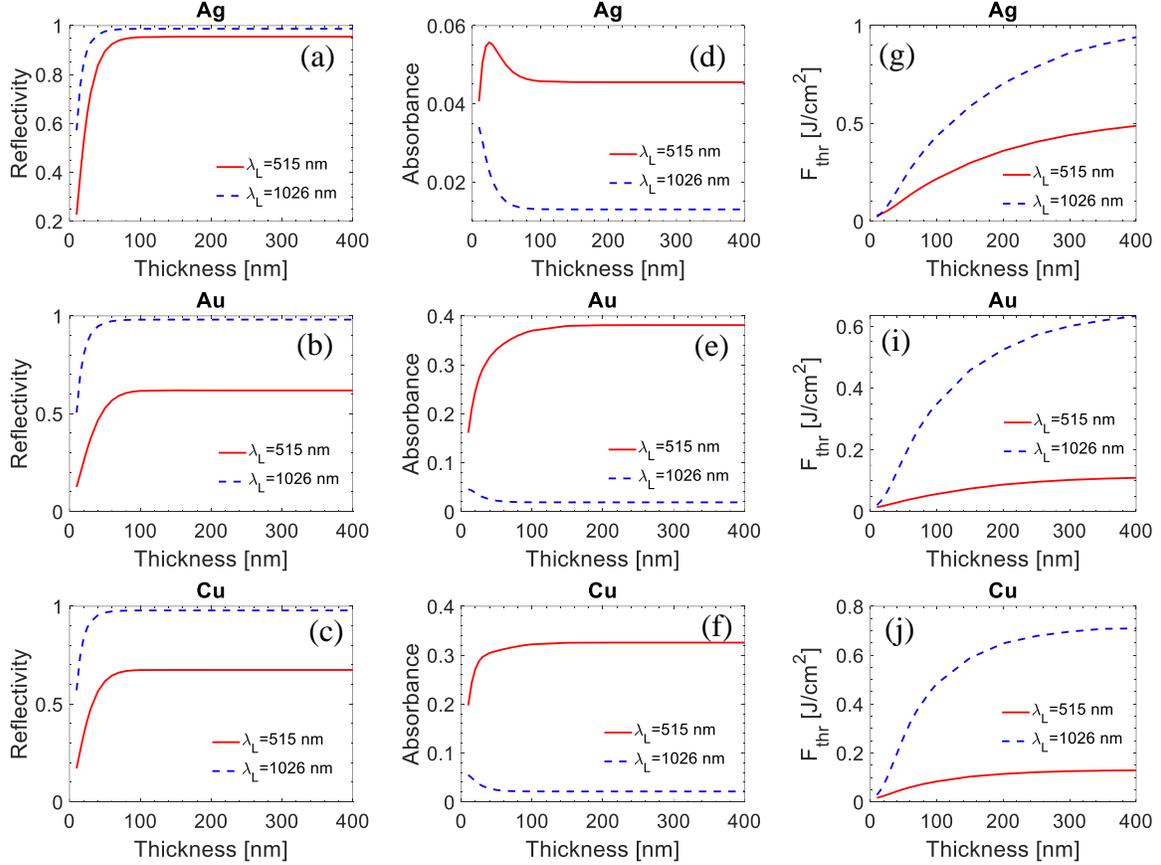

Figure 3: Reflectivity (*left* column), Absorbance (*middle* column) and damage threshold evaluation (*right* column) following irradiation of Au, Ag, Cu with fs laser pulses of two different photon energies ($\lambda_L$=515 nm and $\lambda_L$=1026 nm).

To avoid confusion, it is noted that the reflectivity and absorbance values illustrated in Figs.3-5 correspond to the static values at 300 K. However, a transient variation is considered based on the discussion in Section II to evaluate the thermal response of the electron-phonon systems. The theoretical calculations show that the optical properties of the thin film approximate those of the bulk material when the metal thickness is larger than 80 nm. Interestingly, both the reflectivity and absorptance for some metals (i.e. Stainless Steel, Ti, Ni and Cr) demonstrate a small drop from the peak value before they reach a constant evolution at larger thicknesses. This particular behaviour requires more investigation while a potential relevance of the feature with the electronic structure of those materials should be further explored.

An important issue that requires special focus is whether the remarkable deviation of the calculated reflectivity from the standard values measured or predicted for bulk materials (also, confirmed experimentally for various metals at different wavelengths [42]) is expected to affect the thermal response of the material. To investigate the impact of thickness and the produced absorbed energy on the thermal effects, the theoretical framework presented in Section II is employed to describe the electron dynamics and relaxation processes. The use of the multiscale physical model is aimed to correlate thermal effects with the onset of the material damage. Given that a melting-point-based thermal criterion is chosen to describe the material damage, the objective of the work is the calculation of the minimum (peak) laser fluence that leads to a minimum deformation of the material (i.e. mass displacement).

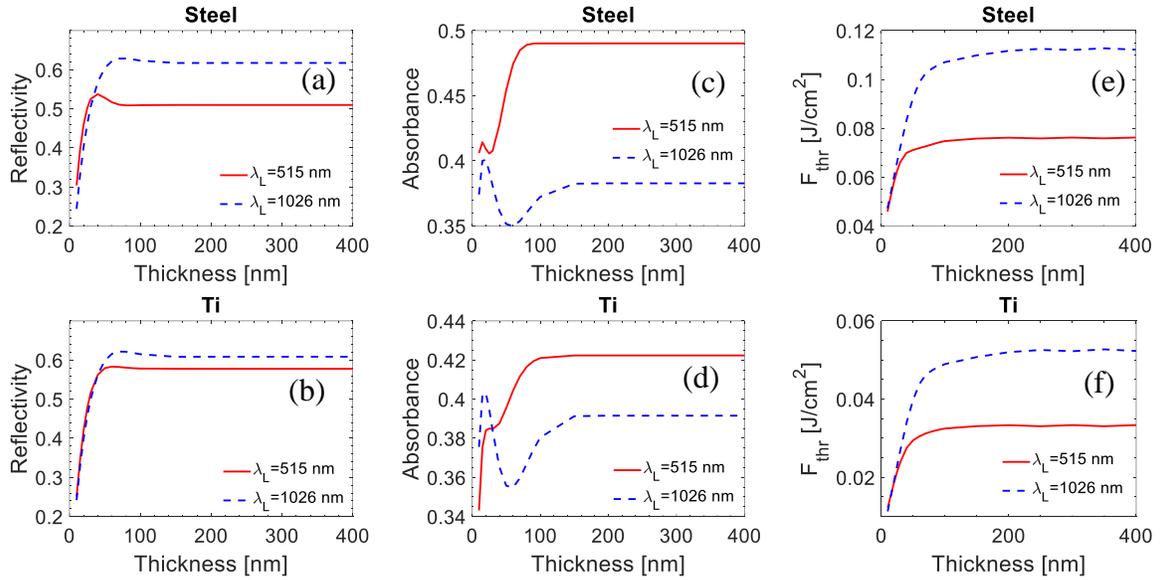

Figure 4: Reflectivity (*left* column), Absorbance (*middle* column) and damage threshold evaluation (*right* column) following irradiation of Steel, Ti with fs laser pulses of two different photon energies ($\lambda_L$=515 nm and $\lambda_L$=1026 nm).

Simulations demonstrate an increase of the damage threshold, $F_{thr}$, with increasing thickness for all materials (*right* column in Figs.3-5) which is justified from the decrease of the produced lattice temperature on the surface of the material due to, primarily, electron diffusion and ballistic transport (thus, less energetic electrons remain on the surface to interact with the phonon system). This is in agreement with the discussion in the beginning of this section and the fact that electron diffusion is facilitated if the metal thickness increases. On the other hand, results show that there is a threshold value for the thickness $d_{thr}$ for most metals after which the material behaves more as a bulk solid, a saturation is reached and, therefore, further increase of $d_{thr}$ does not influence $F_{thr}$. Thus, for larger thicknesses, $F_{thr}$ exhibits an asymptotic behaviour close to the damage threshold of the bulk material. Experimental data in previous reports confirm also this behaviour [41, 43, 45]. To highlight the opto-thermal properties for thicknesses $d$<100 nm, the reflectivity, absorbance and damage threshold variation as a function of the metal thickness, are also illustrated for 10 nm $\leq d \leq$ 100 nm. Results are shown in the Supplementary Material.

Interestingly, the calculations show that for Au, Cu, Ag, Al, the saturation value is delayed while the slow evolution can be attributed to both electron diffusion and the ballistic transport of the hot electrons. Thus, our results show that for most materials, the damage threshold has a linear dependence on film thickness to up to the electron diffusion length $L_{diff}$ which is $L_{diff}$ ~400 nm for Au, Cu, Ag, $L_{diff}$~140 nm for and $L_{diff}$ ~50-60 nm for Stainless Steel, Ni, Ti, Cr (Figs.3-5). A theoretical calculation of the diffusion lengths in previous studies [43] appear to be in good agreement with our simulation results for the damage threshold.

Our simulations for the reflectivity and damage threshold dependence on the metal thickness reveal that there is a discrepancy between the optical and thermal criterion of the definition of a 'bulk' material. In the former case, it is related with the minimum thickness of the metal for which transmissivity becomes very small. By contrast, the thermal criterion suggests a higher value of the $d_{thr}$ at which the material can be described as a bulk metal.

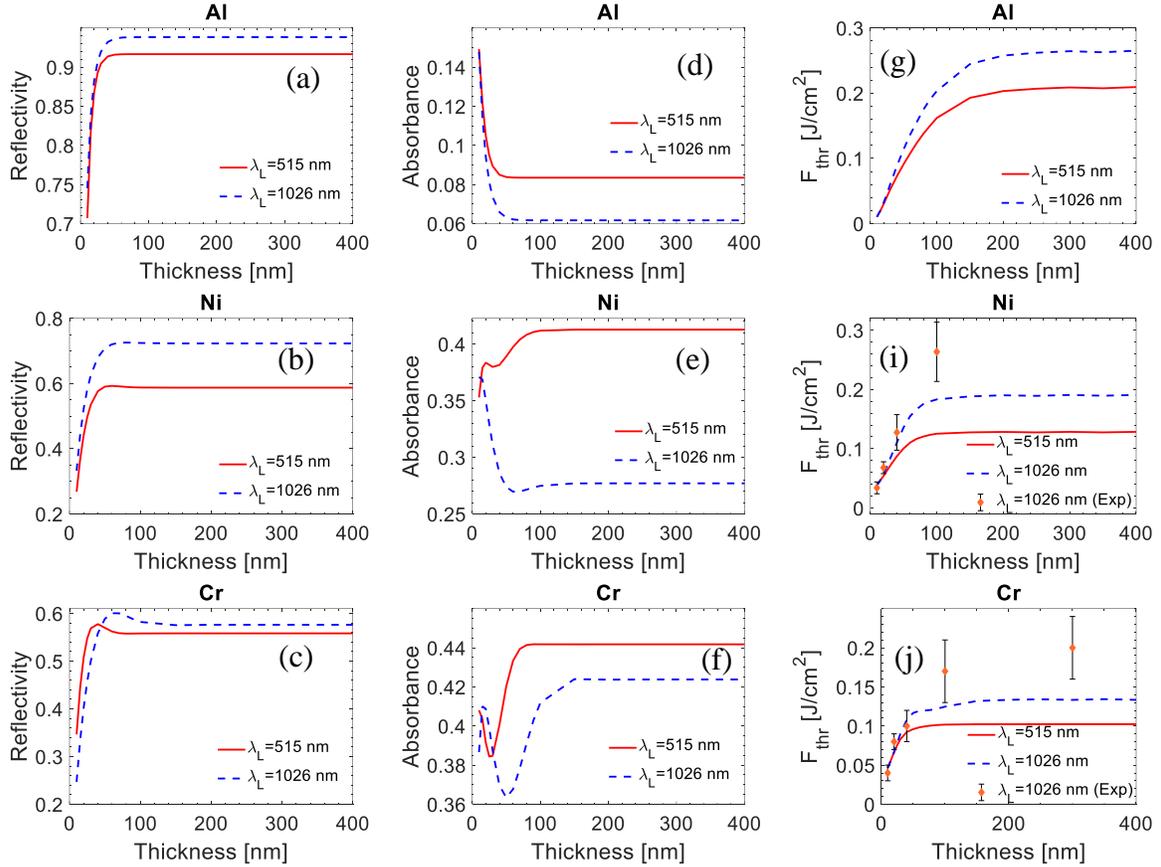

Figure 5: Reflectivity (*left* column), Absorbance (*middle* column) and damage threshold evaluation (*right* column) following irradiation of Al, Ni, Cr with fs laser pulses of two different photon energies ($\lambda_L$=515 nm and $\lambda_L$=1026 nm).

The theoretical predictions for the damage threshold presented in this work appear to agree adequately with experimental data in previous reports for irradiation in relatively similar conditions (for Au [41, 45] (see Fig.6 at $\lambda_L$=400 nm and $\tau_p$=200 fs) or for Ni [41]). To validate the theoretical framework, an analysis of experimental data obtained in this study has also been performed following irradiation of Ni (Fig.5i) and Cr (Fig.5j) targets of various thicknesses at $\lambda_L$=1026 nm at $\tau_p$=170 fs.

For the calculation of single shot damage threshold for Ni and Cr thin films that we used in our experiments, an analysis based in a technique described in Ref. [63] was followed. More specifically, the samples were irradiated with a nearly circular beam of waist radii equal to $w_x = 28.5$ μm, and $w_y = 28.9$ μm for various fluence values and the radius of the damaged area was measured. The error bars were calculated through a standard error propagation methodology, by measuring several quantities during experiments, such as average beam power, beam waist radii, etc. (see Supplementary Material).

As expected, the radius of the damaged area decreases at decreasing fluence and the damage threshold, $F_{thr}$, is defined to be the minimum fluence before the radius of the damaged area vanishes. To calculate the peak energy fluence on the sample, the average power of the beam was measured using a power meter placed after the final focusing lens; it is noted that, losses from the fused silica glass plate of the vacuum chamber were estimated to be approximately ~6.2%. Figs. 7a,b depict SEM images of samples of Cr and Ni, respectively, showing the circular laser processed area. It is noted that the outlines in the samples (Fig.7) have resulted from the removal of a thin oxidized layer due to laser irradiation.

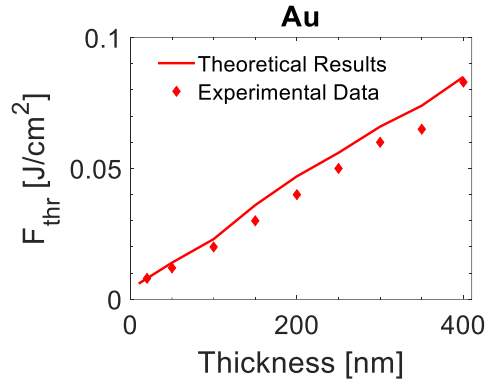

Figure 6: Damage threshold evaluation and comparison with experimental results [41] ($\lambda_L$=400 nm, $\tau_p$=200 fs).

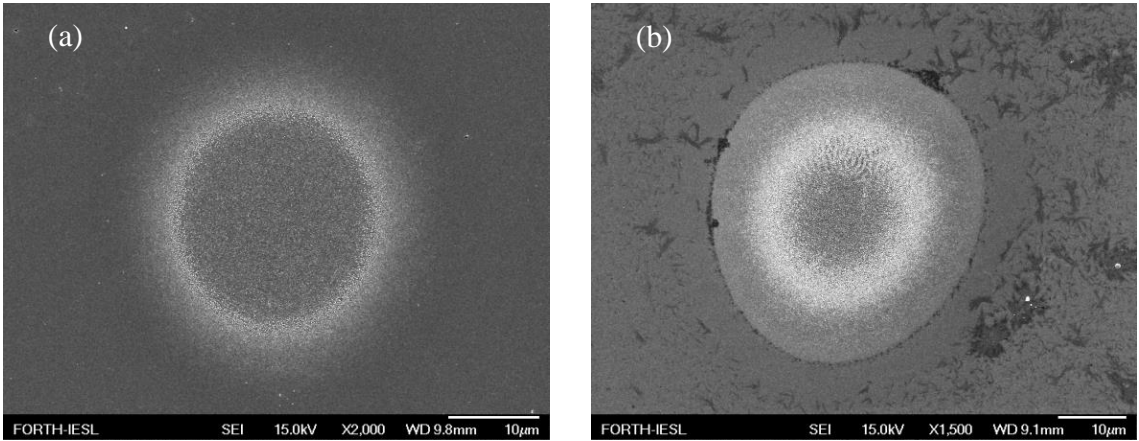

Figure 7: (a) SEM image of a Cr thin film of thickness $d$=100 nm irradiated with a pulse of peak energy fluence $F = 0.27$ J/cm² resulting in a circular damaged area of radius $r_D = 22.5$ μm. (b) SEM image of a Ni thin film of thickness $d$=100nm irradiated with a pulse of peak energy fluence $F = 0.47$ J/cm² resulting in a circular damaged area of radius $r_D = 26$ μm.

It is shown that the theoretical predictions yield a good agreement with the experimental measurements (Figs.5i,j), especially, for small thicknesses. The underestimation for larger thicknesses can be attributed to the fact that the damage threshold criterion used in this report assumed that an onset of material damage occurs even if the minimum volume (one pixel of thickness equal to 0.4 nm according to the discretization) exceeds the melting point of the material. As this size is very small to lead to a sufficiently large molten volume that can produce stable (and visible) material deformation, a larger fluid volume requires an increase of the damage threshold. This implies that an increase in $F_{thr}$ would provide a better agreement with experimental results. By contrast, for small thicknesses, this discrepancy is not evident due to the fact that the electron diffusion is less pronounced which leads to smaller electron (and lattice) temperature *gradients* inside the volume. Therefore, larger volumes exist at a relatively similar temperature, which indicates that for smaller thicknesses it is easier to induce sufficient molten volume that will lead to surface damage. Nevertheless, and in order to be consistent with the widely used definition of the 'melting-point-based' damage threshold (which is the minimum value of the peak fluence that yields lattice temperature above the melting point, regardless of how small the molten volume is), a more detailed analysis of the conditions that lead to the production of larger molten volumes is beyond the scope of this work.

The emphasis in this work was made on the evaluation of the laser conditions that induce minimum damage which is associated with a phase transformation (i.e. melting of the material). Our experimental protocols and results as well as data from previous reports [41-45] were used to validate our model. Certainly, there is a large number of other reports in literature [52, 64-69], however, most of them focus on ablation conditions and the evaluation of ablation threshold.

One aspect that needs to be discussed is the accuracy of the dielectric function expressions that were used for the irradiated solid. As pointed out in the description of the theoretical framework, a Drude-Lorentz model was used to obtain the dielectric function for each metal based on the approach by Rakic *et* al. [56]. In that analysis, the oscillator lengths and spectral widths for the Lorentzian terms that are used were obtained through fitting with experimental data and it was assumed that they do not vary during excitation conditions and at higher electron temperatures. On the other hand, the transient character of the dielectric function was introduced through the inclusion of the relaxation time $\left[A_e \left(T_e^{(m)}\right)^2 + B_e T_L^{(m)}\right]^{-1}$ [57]. This is an approximate method to present a dynamic behaviour into the optical properties of the material. Our simulations showed that the assumption yields results for the damage threshold that agree with the experimental observations. Alternative approaches to evaluate a temporal variation of the dielectric parameter through the employment of a random-phase approximation to compute the electron relaxation time have also been presented in other reports [70-72]. Nevertheless, a more precise investigation would require the employment of more rigorous approaches that reveal not only a time dependent variation but also an electron temperature dependence of the dielectric properties. Such approaches have been developed that were based on the use of first principles and Density Functional Theories to describe the ultrafast dynamics of various materials ranging from metals [18] to semiconductors [19]. These theories could be incorporated into a future and more comprehensive revised model. It is evident that a systematic evaluation of the temporal evolution of the optical parameters should be complemented with the development and use of suitable pump-probe protocols. Appropriate experimental set ups will not only be employed to validate the theoretical models for the calculation of the transient optical parameters but also, would allow a precise determination of the thickness value above which the film would be characterised as bulk (i.e. exhibiting very small transmittance).

Another parameter in the model that can influence the thermal response of the system is the ballistic length. As noted in Section II, $L_b$ is laser wavelength dependent, and therefore, more investigation is required to specify a more precise value for irradiation with $\lambda_L$=1026 nm. Thus, a revised and more consistent model could include a photon energy dependent value of the laser wavelength. Furthermore, the need for further revision to the form of Eq.2 at different laser wavelengths might also be investigated.

With respect to the need to incorporate the impact of the thermal resistance into the model, the damage threshold simulations show that for the three materials for which experimental observations were used, the current framework appears to describe sufficiently the thermophysical behaviour (i.e. for small thicknesses where the EM-PD coupling is more important). Nevertheless, more investigation is required in other laser conditions (i.e. photon energy, pulse durations) or other materials and comparison with relevant experimental protocols to evaluate the contribution of the thermal resistance. Such an approach is expected to identify the regime at which a revision of the model with the incorporation of the thermal resistance contribution is required.

The emphasis of the current work was on the role of the thickness of an irradiated metal that lies on top of a substrate. Both experimental and theoretical results demonstrated the impact of the thickness both on the optical properties and the damage threshold. The theoretical framework can be generalised for more intricate and interesting (from the point of view of applications) systems where multiple layers are present in which the thermophysical and optical properties of the constituent layers are expected to influence the thermal response of the complex [51, 73, 74]. Furthermore, our results show that the enhanced localisation of the energy transferred into the material (for small metal thicknesses), offers a wealth of potential opportunities of nano-patterning.

## V. CONCLUSIONS

A detailed theoretical framework was presented to correlate the impact of the optical properties variation, the damage threshold and the thickness of the most widely used metals for two laser wavelengths of ultrashort laser pulses. It was shown that especially for small thicknesses, the calculated absorbed energy is influenced significantly from the thickness of the irradiated solid, which is reflected on the thermal response of the material. Simulation results which were validated with experimental data on Ni, Cr and Au demonstrated a linear dependence of the damage threshold for thicknesses up to about the optical penetration depth, while at larger thicknesses an asymptotic behaviour close to the damage threshold of the bulk material is followed. The presented model is aimed to provide a tool for an accurate determination of the damage threshold of metals, which is important for a plethora of laser manufacturing approaches.

**ACKNOWLEDGEMENTS**

The authors acknowledge support by the European Union's Horizon 2020 research and innovation program through the project *BioCombs4Nanofibres* (grant agreement No. 862016). G.D.T and E.S. acknowledge funding from *HELLAS-CH* project (MIS 5002735), implemented under the "Action for Strengthening Research and Innovation Infrastructures" funded by the Operational Programme "Competitiveness, Entrepreneurship and Innovation" and co-financed by Greece and the EU (European Regional Development Fund) while G.D.T acknowledges financial support from COST Action *TUMIEE* (supported by COST-European Cooperation in Science and Technology).

**REFERENCES**

[1] E. Stratakis *et al.*, Materials Science and Engineering: R: Reports **141**, 100562 (2020).
[2] A. Y. Vorobyev and C. Guo, Laser & Photonics Reviews **7**, 385 (2012).
[3] V. Zorba, E. Stratakis, M. Barberoglou, E. Spanakis, P. Tzanetakis, S. H. Anastasiadis, and C. Fotakis, Advanced Materials **20**, 4049 (2008).
[4] V. Zorba, L. Persano, D. Pisignano, A. Athanassiou, E. Stratakis, R. Cingolani, P. Tzanetakis, and C. Fotakis, Nanotechnology **17**, 3234 (2006).
[5] J.-C. Diels and W. Rudolph, *Ultrashort laser pulse phenomena : fundamentals, techniques, and applications on a femtosecond time scale* (Elsevier / Academic Press, Amsterdam ; Boston, 2006), 2nd edn.
[6] E. L. Papadopoulou, A. Samara, M. Barberoglou, A. Manousaki, S. N. Pagakis, E. Anastasiadou, C. Fotakis, and E. Stratakis, Tissue Eng Part C-Me **16**, 497 (2010).
[7] A. Papadopoulos, E. Skoulas, A. Mimidis, G. Perrakis, G. Kenanakis, G. D. Tsibidis, and S. E., Advanced Materials **31**, 1901123 (2019).
[8] G. D. Tsibidis, M. Barberoglou, P. A. Loukakos, E. Stratakis, and C. Fotakis, Physical Review B **86**, 115316 (2012).
[9] T. J. Y. Derrien, T. E. Itina, R. Torres, T. Sarnet, and M. Sentis, Journal of Applied Physics **114**, 083104 (2013).
[10] J. Z. P. Skolski, G. R. B. E. Romer, J. V. Obona, V. Ocelik, A. J. H. in 't Veld, and J. T. M. De Hosson, Physical Review B **85**, 075320 (2012).
[11] J. Bonse, J. Krüger, S. Höhm, and A. Rosenfeld, Journal of Laser Applications **24**, 042006 (2012).
[12] F. Garrelie, J. P. Colombier, F. Pigeon, S. Tonchev, N. Faure, M. Bounhalli, S. Reynaud, and O. Parriaux, Optics Express **19**, 9035 (2011).
[13] M. Huang, F. L. Zhao, Y. Cheng, N. S. Xu, and Z. Z. Xu, ACS Nano **3**, 4062 (2009).
[14] J. Bonse and J. Kruger, Journal of Applied Physics **108**, 034903 (2010).
[15] Y. Shimotsuma, P. G. Kazansky, J. R. Qiu, and K. Hirao, Physical Review Letters **91**, 247405 (2003).
[16] B. N. Chichkov, C. Momma, S. Nolte, F. vonAlvensleben, and A. Tunnermann, Applied Physics a-Materials Science & Processing **63**, 109 (1996).
[17] B. Rethfeld, M. E. Garcia, D. S. Ivanov, and S. Anisimov, Journal of Physics D: Applied Physics **50**, 193001 (2017).
[18] E. Bevillon, R. Stoian, and J. P. Colombier, Journal of Physics-Condensed Matter **30**, 385401 (2018).
[19] G. D. Tsibidis, L. Mouchliadis, M. Pedio, and E. Stratakis, Physical Review B **101**, 075207 (2020).
[20] R. Böhme, S. Pissadakis, D. Ruthe, and K. Zimmer, Applied Physics A-Materials Science & Processing **85**, 75 (2006).
[21] B. Chimier, O. Utéza, N. Sanner, M. Sentis, T. Itina, P. Lassonde, F. Légaré, F. Vidal, and J. C. Kieffer, Physical Review B **84**, 094104 (2011).
[22] Y. Jee, M. F. Becker, and R. M. Walser, Journal of the Optical Society of America B-Optical Physics **5**, 648 (1988).
[23] J. Bonse, S. Baudach, J. Kruger, W. Kautek, and M. Lenzner, Applied Physics a-Materials Science & Processing **74**, 19 (2002).
[24] S. Baudach, J. Bonse, J. Kruger, and W. Kautek, Applied Surface Science **154**, 555 (2000).
[25] S. H. Kim, I. B. Sohn, and S. Jeong, Applied Surface Science **255**, 9717 (2009).
[26] G. D. Tsibidis and E. Stratakis, Sci Rep-Uk **10**, 8675 (2020).
[27] I. M. Burakov, N. M. Bulgakova, R. Stoian, A. Rosenfeld, and I. V. Hertel, Applied Physics a-Materials Science & Processing **81**, 1639 (2005).
[28] S. I. Anisimov, B. L. Kapeliovich, and T. L. Perel'man, Zhurnal Eksperimentalnoi Teor. Fiz. **66**, 776 (1974 [Sov. Phys. Tech. Phys. 11, 945 (1967)]).
[29] D. S. Ivanov and L. V. Zhigilei, Physical Review B **68**, 064114 (2003).
[30] R. Shwetharani, H. R. Chandan, M. Sakar, G. R. Balakrishna, K. R. Reddy, and A. V. Raghu, Int J Hydrogen Energ **45**, 18289 (2020).
[31] A. Piegari and F. o. Flory, *Optical thin films and coatings : from materials to applications*, Woodhead publishing series in electronic and optical materials, number 49.
[32] K. V. Sreekanth, S. Sreejith, S. Han, A. Mishra, X. X. Chen, H. D. Sun, C. T. Lim, and R. Singh, Nat Commun **9**, 369 (2018).
[33] M. Kumar, M. A. Khan, and H. A. Arafat, Acs Omega **5**, 3792 (2020).

[34] H. Kim, J. Proell, R. Kohler, W. Pfleging, and A. Pique, Journal of Laser Micro Nanoengineering **7**, 320 (2012).
[35] J. Proll, R. Kohler, A. Mangang, S. Ulrich, M. Bruns, H. J. Seifert, and W. Pfleging, Applied Surface Science **258**, 5146 (2012).
[36] J. Proell, R. Kohler, A. Mangang, S. Ulrich, C. Ziebert, and W. Pfleging, Journal of Laser Micro Nanoengineering **7**, 97 (2012).
[37] W. Pfleging, R. Kohler, M. Torge, V. Trouillet, F. Danneil, and M. Stuber, Applied Surface Science **257**, 7907 (2011).
[38] R. Kohler, H. Besser, M. Hagen, J. Ye, C. Ziebert, S. Ulrich, J. Proell, and W. Pfleging, Microsystem Technologies-Micro-and Nanosystems-Information Storage and Processing Systems **17**, 225 (2011).
[39] R. Kohler, P. Smyrek, S. Ulrich, M. Bruns, V. Trouillet, and W. Pfleging, Journal of Optoelectronics and Advanced Materials **12**, 547 (2010).
[40] E. Skoulas, A. C. Tasolamprou, G. Kenanakis, and E. Stratakis, Applied Surface Science **541** (2021).
[41] S. S. Wellershoff, J. Hohlfeld, J. Güdde, and E. Matthias, Applied Physics A **69**, S99 (1999).
[42] J. Hohlfeld, J. G. Muller, S. S. Wellershoff, and E. Matthias, Applied Physics B-Lasers and Optics **64**, 387 (1997).
[43] J. Güdde, J. Hohlfeld, J. G. Müller, and E. Matthias, Applied Surface Science **127**, 40 (1998).
[44] J. Hohlfeld, S. S. Wellershoff, J. Güdde, U. Conrad, V. Jahnke, and E. Matthias, Chemical Physics **251**, 237 (2000).
[45] B. C. Stuart, M. D. Feit, S. Herman, A. M. Rubenchik, B. W. Shore, and M. D. Perry, J. Opt. Soc. Am. B **13**, 459 (1996).
[46] M. Bonn, D. N. Denzler, S. Funk, M. Wolf, S. S. Wellershoff, and J. Hohlfeld, Physical Review B **61**, 1101 (2000).
[47] Z. Lin, L. V. Zhigilei, and V. Celli, Physical Review B **77**, 075133 (2008).
[48] M. Born and E. Wolf, *Principles of optics : electromagnetic theory of propagation, interference and diffraction of light* (Cambridge University Press, Cambridge ; New York, 1999), 7th expanded edn.
[49] A. Rämer, O. Osmani, and B. Rethfeld, Journal of Applied Physics **116**, 053508 (2014).
[50] J. K. Chen and J. E. Beraun, J Opt a-Pure Appl Op **5**, 168 (2003).
[51] B. Gaković, G. D. Tsibidis, E. Skoulas, S. M. Petrović, B. Vasić, and E. Stratakis, Journal of Applied Physics **122**, 223106 (2017).
[52] A. M. Chen, L. Z. Sui, Y. Shi, Y. F. Jiang, D. P. Yang, H. Liu, M. X. Jin, and D. J. Ding, Thin Solid Films **529**, 209 (2013).
[53] A. M. Chen, H. F. Xu, Y. F. Jiang, L. Z. Sui, D. J. Ding, H. Liu, and M. X. Jin, Applied Surface Science **257**, 1678 (2010).
[54] J. Byskov-Nielsen, J. M. Savolainen, M. S. Christensen, and P. Balling, Applied Physics a-Materials Science & Processing **103**, 447 (2011).
[55] M. Rubin, Sol Energ Mater **12**, 275 (1985).
[56] A. D. Rakic, A. B. Djurisic, J. M. Elazar, and M. L. Majewski, Applied Optics **37**, 5271 (1998).
[57] G. D. Tsibidis, A. Mimidis, E. Skoulas, S. V. Kirner, J. Krüger, J. Bonse, and E. Stratakis, Applied Physics A **124**, 27 (2017).
[58] F. Fraggelakis, G. D. Tsibidis, and E. Stratakis, Physical Review B **103**, 054105 (2021).
[59] B. H. Christensen, K. Vestentoft, and P. Balling, Applied Surface Science **253**, 6347 (2007).
[60] S. Petrovic, G. D. Tsibidis, A. Kovacevic, N. Bozinovic, D. Perusko, A. Mimidis, A. Manousaki, and E. Stratakis, Eur Phys J D **75**, 304 (2021).
[61] D. R. Lide, (CRC Handbook of Chemistry and Physics, 84th Edition, 2003-2004).
[62] L. Guo, S. L. Hodson, T. S. Fisher, and X. F. Xu, Journal of Heat Transfer-Transactions of the Asme **134** (2012).
[63] J. M. Liu, Optics Letters **7**, 196 (1982).
[64] S. Nolte, C. Momma, H. Jacobs, A. Tunnermann, B. N. Chichkov, B. Wellegehausen, and H. Welling, Journal of the Optical Society of America B-Optical Physics **14**, 2716 (1997).
[65] B. Rethfeld, A. Kaiser, M. Vicanek, and G. Simon, Physical Review B **65**, 214303 (2002).
[66] S. Xu, R. Ding, C. Yao, H. Liu, Y. Wan, J. Wang, Y. Ye, and X. Yuan, Applied Physics A **124**, 310 (2018).
[67] X. Jia and X. Zhao, Applied Surface Science **463**, 781 (2019).

[68] A. Nastulyavichus, S. Kudryashov, E. Tolordava, A. Rudenko, D. Kirilenko, S. Gonchukov, A. Ionin, and Y. Yushina, Laser Physics Letters **19** (2022).
[69] A. Nastulyavichus, S. Kudryashov, A. Ionin, and S. Gonchukov, Laser Physics Letters **19** (2022).
[70] G. D. Tsibidis, Journal of Applied Physics **123**, 085903 (2018).
[71] P. A. Danilov, S. I. Kudryashov, K. P. Migdal, A. S. Rivnyuk, and A. A. Ionin, Jetp Lett+ **113**, 297 (2021).
[72] S. G. Bezhanov, P. A. Danilov, A. V. Klekovkin, S. I. Kudryashov, A. A. Rudenko, and S. A. Uryupin, Applied Physics Letters **112** (2018).
[73] S. M. Petrovic, B. Gakovic, D. Perusko, E. Stratakis, I. Bogdanovic-Radovic, M. Cekada, C. Fotakis, and B. Jelenkovic, Journal of Applied Physics **114**, 233108 (2013).
[74] O. V. Kuznietsov, G. D. Tsibidis, A. V. Demchishin, A. A. Demchishin, V. Babizhetskyy, I. Saldan, S. Bellucci, and I. Gnilitskyi, Nanomaterials-Basel **11**, 316 (2021).